\documentclass[12pt]{article}
\usepackage{hyperref}

\usepackage{graphicx}
\begin{document}

\title{\begin{flushright} {\footnotesize IFUP-6/03} \end{flushright} Theoretical
developments for low energy experiments with radioactive beams \footnote {Talk
presented at IX Convegno su Problemi di Fisica Nucleare Teorica,
       Cortona (Italy), October 9-12 2002.}}

\author{Angela Bonaccorso \footnote {\uppercase{I}n collaboration with \uppercase {G.
B}lanchon,\uppercase { D.M. B}rink, \uppercase {J. M}argueron, \uppercase {N. V}inh
\uppercase {M}au}\\
 INFN Sez. di Pisa,
and Dip. di Fisica, Universit\`a di Pisa, 56127 Pisa.
 }

\maketitle

\abstract{In this talk I discuss two types of experiments with exotic nuclei 
 which could be performed at the forthcoming Italian INFN   facilities with radioactive
beams.  First I will discuss nuclear and Coulomb breakup
experiments which involve heavy exotic beams and intermediate incident energies, thus
being suited for the LNL-SPES proposed facility.  Then I will discuss  transfer to the
continuum reactions aiming at performing spectroscopy in the continuum of light unbound
nuclei like 
$^{10}$Li. Such reactions are best matched if the incident beam energy is very low, as
it will be at the LNS-EXCYT facility.  }

\section{Introduction}\label{intro}
Nuclei far from  the stability valley are often called "exotic" because they exhibit 
properties rather different from those of nuclei in the rest of the nuclear chart
\cite{eric}. Most of them are neutron rich and unstable  against $\beta$-decay. It is
interesting to study them because they give information on the structure of matter under
extreme conditions and allow to test nuclear models otherwise based only on
properties of stable nuclei.

  Single
particle degrees of freedom dominate both in the structure description as well as in
the reaction studies of  medium-light unstable nuclei.  So far the most
studied cases have been those of  nuclei like $^{6}He$, $^{11}Be$ $^{11}Li$
which exhibit the so called halo. $^{19}C$ is another interesting candidate still
under investigation \cite{je,nak,typ}.  Heavier nuclei like $^{132}Sn$ have also attracted
much attention.
New techniques are needed to study these nuclei, which combine and unify the
traditional treatment of bound and continuum scattering states
\cite{bb}. 
 Therefore, as in the early stages of Nuclear
Physics, research on light exotic nuclei has concentrated on studying
elastic scattering \cite{bc} and spectroscopic properties like the determination of
single particle state energies,  angular momenta and spectroscopic factors
\cite{je,nak}.

\section{Reaction models for structure studies: exclusive and inclusive breakup}

One of the most suited measurement for an exotic projectile is the single-neutron
removal
   cross section, in which only the projectile residue, namely the
   core with one less nucleon, is observed in the final state. This information
together with the calculated cross sections \cite{je,nak}, has been
used to extract single particle spectroscopic factors as in traditional transfer
reactions. Besides
   the integrated removal cross section, denoted by $\sigma_{-n}$, the
   differential momentum distribution $d^3\sigma/d k^3$ is also
   measured. A particularly useful cross section is $d \sigma /d
   k_z$, the removal cross section differential in longitudinal momentum. It has been
used to determine the angular momentum and spin of the neutron initial state \cite{je} in
a way similar to that proposed in \cite{bb,tiina}.
   If the final state neutron can also be measured, the corresponding
   coincident cross section $A_p \rightarrow (A_p-1) +n$ is called the
   diffractive (or elastic ) breakup cross section if the interaction responsible for the
removal is the neutron-target nuclear potential \cite{anne,ab}. In the case of heavy
targets the coincident cross section contains also the contribution from Coulomb breakup
due to the core-target Coulomb potential which acts as an effective force on the
neutron. This observable is very useful to disentangle the reaction mechanism
\cite{jer}. The difference between the removal and  coincident cross sections is called
the stripping (or absorption) cross section.

All theoretical methods used so far rely on a basic approximation to describe the collision with
   only the three-body variables of nucleon coordinate, projectile
   coordinate, and target coordinate. Thus the dynamics is controlled by the
   three potentials describing nucleon-core, nucleon-target, and core-target
   interactions. In most cases the projectile-target relative motion is treated
   semiclassically by using a trajectory of the center of the projectile
   relative to the center of the target ${\bf R}(t)={\bf d}+{\bf v}t$
   with constant velocity $v$ in the $z$ direction and impact parameter
   {\bf d} in the $xy$ plane.

A full description of the treatment of the scattering
equation for a projectile which decays by single neutron breakup following its
interaction with the target, including core recoil, can be found in \cite{bb,jer}.

 In ref.\cite{jer} it was shown that the combined effect of the nuclear and Coulomb
interactions to all orders can be taken into account by using the potential
${V}=V_{nt}+V_{eff}$ sum of the neutron-target optical potential and the Coulomb dipole
potential.   If for the neutron final continuum  wave function we take a distorted
wave of the  eikonal-type, then   the amplitude  
 for a
   transition from a  nucleon bound state $\phi_i$ in the projectile to a final
continuum state
 becomes :
\begin{equation}
A_{fi}\left(  \mathbf{k,}\mathbf{d}\right)  =\frac{1}{i\hbar}\int
d^{3} {\bf r} \int dte^{-i{\bf k \cdot r}+i\omega t} e^{\left(  \frac{1}{i\hbar
}\int_{t}^{\infty}{V}\left(  {\bf r},t^\prime\right)  dt^\prime\right)  }
{V}\left(  {\bf r},t\right)  \phi_{l_im_i}\left(  \mathbf{r}\right)
\label{amp1}%
\end{equation}
where $\omega=\left(  {{\varepsilon_f}^{\prime}}-\varepsilon_{0}\right)  /\hbar
$ and $\varepsilon_0$ is the neutron initial bound state energy  while
${{\varepsilon_f}^{\prime}}$ is the neutron-core final continuum energy. 
Eq.(\ref{amp1}) is appropriate to calculate the coincidence cross sections $A_p
\rightarrow (A_p-1) +n$ discussed in the previous section. Finally the
differential probability with respect to the neutron energy and angles can be
written as
${d^3P_{nc}(d)\over d{{\varepsilon_f}^{\prime}}\sin\theta d\theta d\phi}
={1\over 8\pi^3}{m k_n\over \hbar^2}{1\over
2l_i+1}\Sigma_{m_i}| A_{fi}|^2 ,\label{anc}$
where $A_{fi}$ is given by Eq.(\ref{amp1}) and we have averaged over the neutron
initial state.

   The effects associated with the core-target interaction will be included by 
   multiplying the above probability by $P_{ct}(d)=|S_{ct}(d)|^2$ the
probability for the core to be left in its ground state, defined in
terms of a core-target S-matrix function of  $d$, the core-target distance of
closest approach \cite{ab} . 

  Thus the double differential cross section  is 
    \begin{equation}
     {d^2\sigma\over  {d{{\varepsilon_f}^{\prime}} d\Omega}}=C^2S
    \int_0^{\infty} d{\bf d} {d^2 P_{nc}({\bf k},d)\over
d{{\varepsilon_f}^{\prime}}\Omega} 
    P_{ct}(d), \label{cross} \end{equation}
   and $C^2S$ is the spectroscopic
   factor for the initial single particle orbital.

Inclusive cross sections in which only the core
with $(A_p-1)$ nucleons is detected need to take into account also the
absorption of the neutron by the imaginary part of the n-target optical
potential. For
such reactions the Coulomb recoil effect can be neglected but the distorted eikonal-type wave function used in
Eq.(\ref{amp1}) is not accurate enough, in particular if the final continuum states are single particle
resonances in the target plus one neutron nucleus. Then
 a distorted final neutron wave function, calculated by an optical model  
will be used. Also since the neutron is not detected one integrates over the neutron angles.  Thus, according
to \cite{bb} the final neutron probability energy spectrum with respect to the target reads
\begin{equation}{dP\over d\varepsilon_f} \approx \Sigma_{j_f}(|1-\bar S_{j_f} |^2+1-|\bar
S_{j_f} |^2) (2j_f+1)(1+F_{j_f,j_i})
{C_i^2\over mv^2}{1\over
2k_f} {e^{-2\eta d}\over 2\eta d}M_{l_fl_i},\label{dpde}\end{equation} 
where $\bar S_{j_f}$ is the neutron-target optical model S-matrix,
$F_{j_f,j_i}$ is an $l$ to $j$ coupling factor, $\eta$ is the
transverse component of the neutron momentum which is conserved in the
neutron transition, $d$ is the core-target impact parameter,
$C_i$ is the initial state asymptotic normalization constant and $M_{l_fl_i}$
is a factor depending on the angular parts of the initial and final  wave
functions, $v$ is the relative motion velocity at the distance of closest approach.
\section{Applications}

We are going to discuss now a series of experiments and corresponding theoretical 
calculations aimed at extracting spectroscopic  information on one-neutron and
two-neutron halo nuclei and to determine properties of  neutron-exotic nucleus
interactions.

\subsection{ Neutron differential cross sections  following breakup.}

\begin{figure}[h]
\includegraphics[scale=0.5,angle=90]{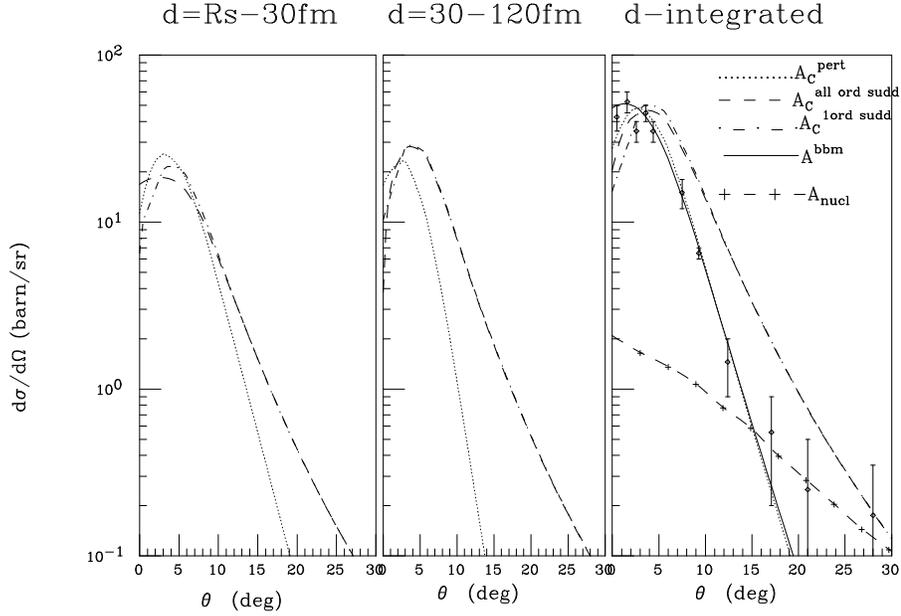}   
\footnotesize\caption{Neutron angular distributions distribution after nuclear-Coulomb
breakup.}\label{eps2}
\end{figure}

$^{11}Be$ is probably the best known one-neutron halo nucleus since experimental information has
been available for long time . The ground state is a $2s_{1/2}$ state with
separation energy of 0.5MeV and spectroscopic factor $C^2S=0.77$ \cite{mill}. Therefore it has been
used as a test case for reaction models which use the above basic structure information as
input \cite{anne,ab,abfc}.

On the other hand a more recent work \cite{jer} has improved  the previous knowledge 
of the breakup reaction, by
studying the Coulomb-nuclear interference effects according to Eq.(\ref{amp1}).
We  report here on new calculations 
with Eq.(\ref{amp1}) to study higher order effects.  Three limits of Eq.(\ref{amp1}) have been tested. The
first is the sudden approximation in which  $\omega=0$ and Eq.(\ref{amp1}) can be
calculated with  nuclear and/or Coulomb to all orders. We call the corresponding amplitudes
$A_{C}^{all~ord~sudd}$ and $A_{nucl}$. Then we have studied the first order
approximation for the Coulomb term in which
$e^{\left( 
\frac{1}{i\hbar }\int_{t}^{\infty}{V_{eff}}\left(  {\bf r},t^\prime\right)  dt^\prime\right) 
}=1$ but the
$\omega t$ term is kept (this is the standard first order perturbation theory
amplitude $A_C^{pert}$) and  finally the sudden approximation restricted to first order  giving
$A_{C}^{1ord~sudd}$. The main results of our new calculations are shown in Fig.(1a) and
(1b) which give the neutron final angular distribution in the laboratory for the
reaction
$^{11}Be(^{197}Au,^{197}Au) ^{10}Be+n$ at 41 A.MeV
\cite{anne}. The curves in Fig.(1a) indicate that for $R_s < d < 30fm $ the results
obtained with $A_{C}^{all~ord~sudd}$ are equal to those obtained with $A_{C}^{1ord~sudd}$,
starting from about $\theta=10deg$ thus showing that
 higher order terms need to be considered only at  small angles. On the
other hand for $d >30fm$ we find that higher order effects are always negligible since
using
$A_{C}^{all~ord~sudd}$ or $A_{C}^{1ord~sudd}$ does not give any difference.  We trust that higher order terms 
are calculated correctly  by the sudden approximation because we  have checked that
the second order term calculated with full time dependence or in the sudden limit, gives
the same results. Then we 
conclude that first order time dependent perturbation theory is valid and appropriate apart from the small
impact parameter, small angle region. Thus in Fig.(1c) we finally give by the dotted curve  the results of
the simple first order perturbation theory  while the solid curve is the all order
calculation according to an amplitude defined as
$A^{'bbm'}=A_{C}^{all~ord~sudd}+A_C^{pert}-A_{C}^{1ord~sudd}$. Such an amplitude is valid at all core-target
impact parameters, contains all order contributions and it does not give rise to any divergence in the
final integral over impact parameters in Eq.(\ref{cross}) because the divergent term $A_{C}^{1ord~sudd}$ is
substituted by $A_C^{pert}$. Data points are from
\cite{anne}. The theoretical calculations have been multiplied by the known spectroscopic
factor. Analysis
of the type presented in this section have been used and could be used in the future to
extract spectroscopic factors. 

\begin{figure}[h]
\includegraphics[scale=0.45,angle=90]{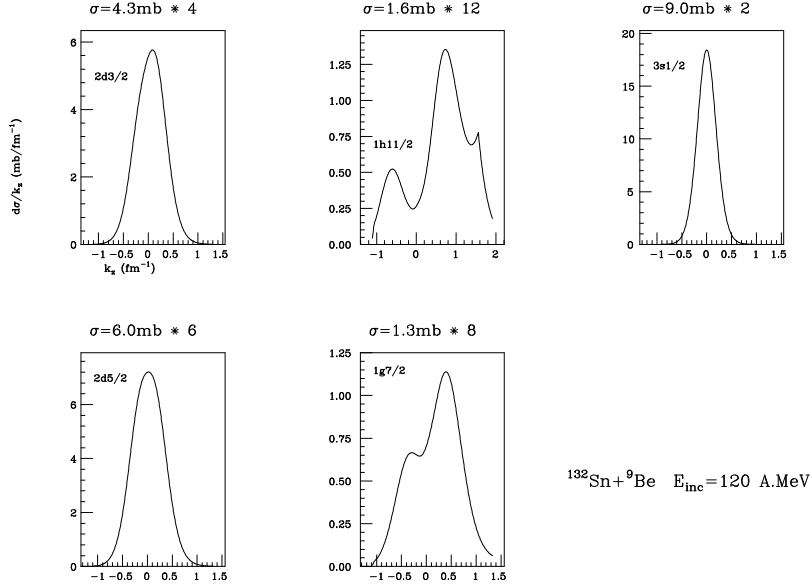}  
\footnotesize\caption{Neutron parallel momentum  spectra following breakup of several single
particle states in
$^{132}$Sn.}\label{eps4}
\end{figure}

Another neutron-rich nucleus which recently has been discussed at length \cite{lev} is $^{132}Sn$ which
should exhibit the N=82 shell closure.
It is expected that the spin-orbit splitting should decrease or even vanish for
such very neutron-rich isotopes. A way to study the spectroscopy of this nucleus would be to measure the
absolute cross sections and core parallel momentum distributions from the breakup of neutrons  of
different orbitals. Such a measurement would be possible by taking coincidences between
the core and $\gamma$-rays as done in \cite {je}. As one can see from Fig.(2) the
shapes of the parallel momentum distributions, calculated according to Eqs.(2) and (3) depend on the initial state
and are determined by the spin-orbit coupling coefficient $F_{j_f,j_i}$ of
Eq.(\ref{dpde}). They reflect the momentum distribution
of the neutron  in the projectile and thus  they  and can be used to determine it. The
ratio of the measured and calculated absolute cross sections on the other hand
determine  the single particle spectroscopic factor according to Eq.(\ref{cross}). We
give the calculated values on top of each initial state spectrum in Fig.(2) together
with the shell model occupation number.

\subsection{$^{10}Li$ spectrum and $^{11}Li$ properties.}
\begin{figure}[h]
\includegraphics[scale=0.5,angle=90]{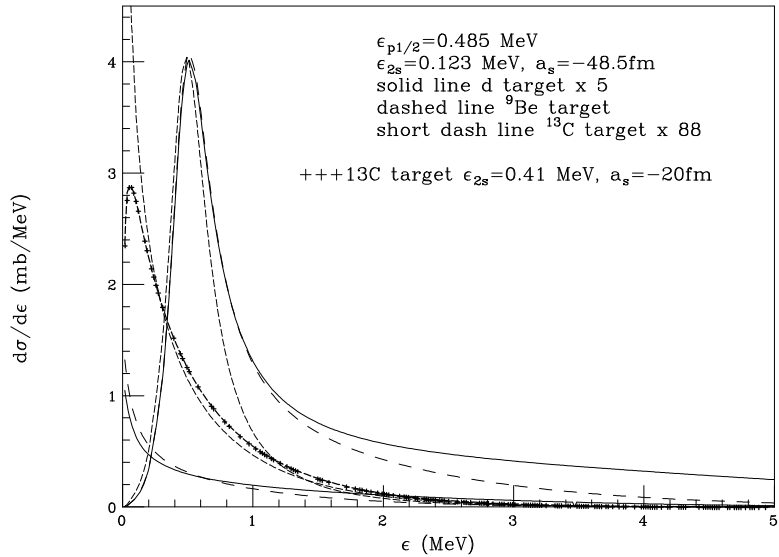}   
{\small \caption{Neutron-$^{9}$Li relative energy spectra for transfer to the s and
p continuum states in $^{10}$Li.}}\label{eps5}
\end{figure}

We discuss now the results of a possible reaction aiming at clarifying the structure
$^{11}Li$ which 
  has been a challenge for long time \cite{menic,tho}-\cite{sf}.
A similar reaction could be performed at the LNS with the $^{8}Li$ beam to study the
structure of the unstable  $^{9}Li$ and to determine the n-$^{8}Li$ interaction. 

$^{11}Li$ and
$^{6}He$ are two-neutron halo nuclei. Their corresponding A-1
systems, such as $^{10}Li$  are unbound and have therefore been
very difficult to study from the experimental point of view \cite{cha}-\cite{sf}. However $^{11}Li$
is bound  thanks to the pairing force acting between the two extra neutrons.
     
 Recently the experiment $d(^{11}Be,^{3}He)^{10}Li$ 
\cite{sf} has  confirmed that the ground state of
$^{10}Li$ is a 2s virtual state. 
 
Tree body models of $^{11}Li$ need as a fundamental ingredient the n-core
(n-$^{9}$Li) interaction, which in turn determines the energies of the low energy unbound
states in $^{10}$Li. Following ref.\cite{pvm} the two neutron hamiltonian is
$H_{2n}=h_{1}+h_{2}+V_{nn}$.
$V_{nn}$ is the zero-range paring interaction. The single
neutron hamiltonian is 
$h=t+ V_{cn}$ where $t$ is the kinetic energy and $V_{cn} =V_{WS}+\delta V$ is the
neutron-core interaction. It is given by the usual Woods-Saxon potential plus
spin-orbit plus a correction
$\delta V$ which originates from particle-vibration couplings.  They are important for low
energy states but can be neglected at higher energies. If Bohr and Mottelson collective
model is used for the transition amplitudes between zero and one phonon states, then 
 $
\delta V(r)=16\alpha {e^{{{2(r-R)}/ a}}/({1+e^{{{(r-R)}/
 a}}})^4 }$\
where $R\approx r_0A^{1/3}$. According to \cite{pvm}  the best parameters
for the n-$^{9}$Li  Woods-Saxon and spin-orbit potential are
$V_0=-39.83 MeV$, $V_{so}=7.07 MeV$, $r_0= 1.27 fm$, $a=0.75 fm$.
The corresponding energies obtained for the 2s and 1p$_{1/2}$ states are given in Table
1, together with the values of the strength $\alpha$ of the correction potential
$\delta V$.

\begin{table}[h]
\footnotesize\caption{Energies of the s and p states, width of the p-state, scattering length
of the s-state and strength of the $\delta V$ potential. (a) bound-state 
calculation, (b) scattering state calculation.}\vskip.3in \begin{center}
{\footnotesize
\begin{tabular}{lccccc}
\hline\
                        & (a) & (b)&$\Gamma$(MeV)&$a_s$(fm)& $\alpha(MeV)$\\
$\epsilon_{2s_{1/2}}(MeV)$       & 0.123 & 0.17 &&-48.5&-13.3\\
&&0.45&&-20&-14.0\\
$\epsilon_{1p_{1/2}}(MeV)$ & 0.485 & 0.595 &0.48&&3.3\\
\hline
\end{tabular}}\end{center}
\end{table}

 It is therefore extremely  important to
determine experimentally the energies of the two unbound $^{10}$Li states such that the interaction
parameters can be deduced. Two $^{9}Li(d,p)^{10}Li$ experiments have recently been
performed, at MSU at 20 A.MeV \cite{santi} and  at the CERN REX-ISOLDE facility
at 2 A.MeV\cite{bj}. For such transfer to the continuum reactions the predictions of the theory underlined in
Sec. 2, Eq.(3) are very accurate. The sensitivity of the results on the target and on the energies assumed
for the s and p states has been studied by
calculating the reaction $^{9}Li(X,X-1)^{10}Li$ at 2
A.MeV for three targets d, $^{9}$Be, $^{13}$C. 
 The  $^{13}$C target has been chosen
because in such a case the neutron transfer to the 2s state in $^{9}$Li would be a
spin-flip transition which as it is well known  are enhanced at low incident
energy. For the other two cases the transfer to the 2s state is a non spin-flip
transition which is hindered.  Fig.(3) shows the neutron energy spectrum relative to 
$^{9}$Li obtained with the interaction and single particle energies of Tables 1. In the
case of the 2s virtual state  also the scattering length 
$a_s=-  \mathrel{\mathop{lim}\limits_{k \to
0\hspace{.28em}}}{tan\delta_0\over k} $ is given.  The
peak of the p-state will determine without ambiguity the energy of the
state in a target independent way. The width  is modified by the
reaction mechanism, but it can  be deduced from the 
 phase shift behavior once that the energy is fixed.  
 There is a larger probability of population of  the s-state in the spin-flip
reaction initiated by the carbon target. A measure of the line-shape (or
spectral function) and absolute value of the cross section will determine the
energy of the state also in this case. The integral over energy of the  distribution  determines
the spectroscopic factor of the state. There is no spreading width of the
single particle state since the n-$^9$Li interaction is real at such low energies,
 the first excited state of $^9$Li being at
$E^*=2.7MeV$. This means also that the "resonances" of the n-$^9$Li system are not compound nucleus
resonances but rather elastic scattering resonances and therefore arise only from  the  term
$|1- \bar S|^2$ in  Eq.(3). The
sensitivity of the model calculation on the energy of the state is shown in Fig.(3) by  results 
indicated with the crosses  obtained when for the the s-state 
$\varepsilon_{2s}=0.45 MeV$ corresponding to $a_s=-20 fm$. A clear peak
appears even if located at very small energy. The appearance of  a peak
 in the transfer spectrum, depends on the  behavior in Eq.(3) of the  term
$|1- \bar S|^2$ which has always a maximum
value equal to 4 at the energy of the state, with respect to the product of the other terms which  have a
divergent-like behavior as the energy approaches zero. We conclude that if a transfer to the continuum 
experiment could measure with sufficient energy resolution the line-shapes or energy distribution
functions for the s and p-states in $^{10}$Li our theory would be able to fix unambiguously the energies of
the states. Those in turn could be used to test microscopic models of the n-$^{9}$Li interaction.

\section{Conclusions and future challenges}

Physics with radioactive beams
is an extremely fascinating field in which the interplay between the understanding of the
nuclear structure and that of the reaction mechanism is very strong and  an enormous
number of progress has been made in the last few years. However a number of
improvements both experimental as well as theoretical need to be pursued. We have
shown that there are  experiments which could be performed at the
forthcoming facilities at the Italian
 Nuclear Physics Laboratories for which the theoretician's community is ready to give
the appropriate support.

\end{document}